\documentstyle[12pt,aaspp4,flushrt]{article}

\slugcomment{Accepted for publication in ApJ}

\newcommand\A{{\cal A}}
\newcommand\al{\alpha}
\newcommand\g{\gamma}
\newcommand\G[2]{\Gamma\left({#1}\,;\,{#2}\right)}
\newcommand\s{\sigma}
\renewcommand\th{\theta}
\newcommand\kms{\mbox{km s$^{-1}$}}
\newcommand\tot{{\rm tot}}
\newcommand\fevol{f_{\rm evol}}
\newcommand\avg[1]{\langle{#1}\rangle}

\newcommand\Lcut{L_{\rm cut}}
\newcommand\faint{{\rm faint}}
\newcommand\bright{{\rm bright}}

\newcommand\refeq[1]{eq.~(\ref{eq:#1})}
\newcommand\refEq[1]{Eq.~(\ref{eq:#1})}

\newcommand\reffig[1]{Figure~\ref{fig:#1}}
\newcommand\reffigs[2]{Figures~\ref{fig:#1} and \ref{fig:#2}}

\newcommand\reftab[1]{Table~\ref{tab:#1}}
\newcommand\reftabs[2]{Tables~\ref{tab:#1} and \ref{tab:#2}}

\begin{document}

\title{What Fraction of Gravitational Lens Galaxies Lie in Groups?}
\author{Charles R.\ Keeton, Daniel Christlein, and
  Ann I.\ Zabludoff}
\affil{Steward Observatory, University of Arizona \\
  933 N.\ Cherry Ave., Tucson, AZ 85716}

\begin{abstract}

We predict how the observed variations in galaxy populations with
environment affect the number and properties of gravitational
lenses in different environments. Two trends dominate: lensing
strongly favors early-type galaxies, which tend to lie in dense
environments, but dense environments tend to have a larger ratio
of dwarf to giant galaxies than the field. The two effects nearly
cancel, and the distribution of environments for lens and non-lens
galaxies are not substantially different (lens galaxies are
slightly less likely than non-lens galaxies to lie in groups
and clusters). We predict that $\sim$20\% of lens galaxies are in
bound groups (defined as systems with a line-of-sight velocity
dispersion $\s$ in the range $200 < \s < 500$ \kms), and another
$\sim$3\% are in rich clusters ($\s > 500$ \kms). Therefore at
least $\sim$25\% of lenses are likely to have environments that
significantly perturb the lensing potential. If such perturbations
do not significantly increase the image separation, we predict that
lenses in groups have a mean image separation that is $\sim\!0\farcs2$
{\it smaller} than that for lenses in the field and estimate that
20--40 lenses in groups are required to test this prediction with
significance. The tail of the distribution of image separations is
already illuminating. Although lensing by galactic potential wells
should rarely produce lenses with image separations $\th \gtrsim
6\arcsec$, two such lenses are seen among 49 known lenses, suggesting
that environmental perturbations of the lensing potential can be
significant. Further comparison of theory and data will offer a
direct probe of the dark halos of galaxies and groups and reveal
the extent to which they affect lensing estimates of cosmological
parameters.

\end{abstract}

\section{Introduction}

Gravitational lensing of background sources by foreground galaxies
offers a powerful probe of galaxy structure and evolution at
intermediate redshifts ($0.3 \lesssim z \lesssim 1$; e.g., Keeton,
Kochanek \& Falco 1998; Kochanek et al.\ 2000a). Lenses are also
used to constrain the cosmological parameters $H_0$, $\Omega_M$,
and $\Omega_\Lambda$ (e.g., Falco, Kochanek \& Mu\~noz 1998; Helbig
et al.\ 1999; Koopmans \& Fassnacht 1999; Witt, Mao \& Keeton 2000;
and references therein). These applications rely on accurate models
of the mass distribution responsible for the lensing, but the
models may be complicated by contributions of the lens galaxy's
environment to the lensing potential, especially when the lens
galaxy lies in a poor group or rich cluster of galaxies.

A crucial step in understanding how environment may affect lensing
constraints on dark matter halos and cosmological parameters is
determining the fraction of lens galaxies in groups and in clusters.
The environments of most lenses are not known. Three lens galaxies
are confirmed group members (PG 1115+080, B 1422+231, and
MG 0751+2716; Kundi\'c et al.\ 1997ab; Tonry 1998; Tonry \& Kochanek
1999), and another four lie in clusters (RXJ 0911+0551, RXJ 0921+4528,
Q 0957+561, and HST 1411+5221; Young et al.\ 1981; Fischer, Schade
\& Barrientos 1998; Kneib, Cohen \& Hjorth 2000; Mu\~noz et al.\
2000).  While environmental effects have been included in models
of some of these lenses (e.g., Schechter et al.\ 1997; Keeton \&
Kochanek 1997; Barkana et al.\ 1999; Bernstein \& Fischer 1999; Chae
1999; Keeton et al.\ 2000a; Leh\'ar et al.\ 2000), lens environments
have not been addressed statistically (see Keeton, Kochanek \& Seljak
1997 for an initial analysis). In particular, previous predictions
of the statistics of galaxy lenses could not examine environmental
effects because they assumed an environment-independent galaxy
luminosity function (e.g., Turner 1980; Turner, Ostriker \& Gott
1984; Fukugita \& Turner 1991; Kochanek 1993a, 1996ab; Maoz \& Rix
1993; Wallington \& Narayan 1993; Falco et al.\ 1998; Quast \& Helbig
1999; Helbig et al.\ 1999).

The existence of lenses in groups and clusters is not surprising.
Lensing selects galaxies by mass and thus preferentially selects
early-type galaxies, which are most common in dense environments
(e.g., the morphology--density relation; Dressler 1980). However,
recent studies suggest that richer environments tend to have a
higher ratio of dwarf to giant galaxies than the field (Bromley et
al.\ 1998b; Christlein 2000; Zabludoff \& Mulchaey 2000), and dwarf
galaxies are much poorer lenses than giant galaxies. The
morphology--density relation increases the chance of having a lens
in a group or cluster while the dwarf-to-giant ratio decreases it,
so the distribution of lens galaxy environments depends on how
these two effects combine quantitatively.

In this paper, we examine how changes in the type and luminosity
distribution of galaxies with environment (``population
variations'') affect lens statistics. Using new results that
quantify the type and environment dependence of the galaxy
luminosity function (Bromley et al.\ 1998ab; Christlein 2000), we
obtain the first predictions of the distribution of lens galaxy
environments and of statistical differences in the properties of
lenses in different environments. We neglect contributions to
the lensing potential from other galaxies and the extended dark
halo of the group or cluster (``potential perturbations''), which
introduce a tidal shear that distorts image configurations and a
gravitational focusing that increases image separations (e.g.,
Schneider, Ehlers \& Falco 1992). This assumption not only
simplifies the analysis, but also constitutes the null hypothesis
as to whether potential perturbations contribute (statistically)
to lensing.

The organization of the paper is as follows. In \S 2 we discuss the
components of the statistical calculations, including galaxy
luminosity functions and the gravitational lens model. In \S 3 we
use simple analytic results to highlight the trends of
environmental effects in lens statistics. In \S 4 we use empirical
galaxy luminosity functions to obtain quantitative predictions of
the environmental effects. Finally, in \S 5 we offer a summary and
discussion.

\section{Data and Methods}

\subsection{Galaxy populations}

We must formally describe a population of galaxies that can act as
gravitational lens galaxies in order to compute the set of lenses
they can produce. Depending on details of the observational sample,
galaxies may be binned by type, environment, and/or redshift.
Within each discrete bin, the distribution of galaxy luminosities
is described using a smooth galaxy luminosity function (GLF)
$\phi(L)$ such that $\phi(L)\,dL$ is the comoving number density of
galaxies with luminosity between $L$ and $L+dL$. The GLF is usually
parameterized as a Schechter (1976) function with the form
\begin{equation}
  \phi_i(L) = {n^*_i \over L^*_i}\, \left({L \over L^*_i}\right)^{\al_i}
    e^{-L/L^*_i} , \label{eq:schech}
\end{equation}
where $i$ labels the type, environment, and/or redshift bin. The
GLF is described by a characteristic comoving number density
$n^*_i$, a characteristic luminosity $L^*_i$, and a faint end slope
$\al_i$. The differences between GLFs in different bins are
characterized by differences in the Schechter parameters $n^*_i$,
$L^*_i$, and $\al_i$.

Current surveys are not large enough to bin galaxies by type,
environment, and redshift simultaneously; only recently has it
become possible to examine even two of the three quantities.
Because we are interested in the effects of environment, we adopt
the sample of galaxies from the Las Campanas Redshift Survey (LCRS,
Shectman et al.\ 1996). This survey provides a large-volume, R-band
selected sample of the nearby universe that offers four benefits
for our study. First, it samples down to an estimated completeness
limit of $M_R = -17.5 + 5 \log h$, or nearly three magnitudes
below $M^*$. Second, its large volume offers a fair sample of
environments from the field to groups and clusters. Third, its
large number of galaxies means that galaxies can be binned by both
type and environment with sufficient statistics to discriminate
between GLFs in different bins.

The fourth benefit is that there are two independent and
complementary analyses of type and environmental dependences in
the LCRS GLFs. Bromley et al.\ (1998ab, hereafter B98) classify
galaxies using six type and two environment classes. They define
six spectral types (which they call ``clans'') derived from a
principal component analysis of important spectral features. The
clans smoothly span the range from quiescent galaxies that have
substantial absorption lines, 4000 \AA\ breaks, and old stellar
populations (Clans 1 and 2) to star forming galaxies with prominent
emission lines and a significant fraction of young stars (Clans 5
and 6). B98 define two environment categories based on the local
3-d number density of galaxies: ``high density'' environments
correspond roughly to groups and clusters of galaxies (identified
using a friends-of-friends algorithm), while ``low density''
environments include only galaxies that lie outside groups.
\reftab{B98lf} gives the parameters for the GLFs derived by B98.

The study by Christlein (2000, hereafter C00) uses a different
scheme for classifying type and environment, placing more emphasis
on environment than type. C00 defines two galaxy types, using the
equivalent width of {\sc [Oii]} $\lambda$3727 \AA\ to separate
``emission line'' ($\ge$5 \AA; EL) from ``non-emission line''
($<$5 \AA; NEL) galaxies. He defines environments by identifying
groups (using a friends-of-friends algorithm) and classifying
galaxies by the velocity dispersion of the group in which they
reside. \reftab{C00lf} gives the parameters for the GLFs derived
by C00.

B98 and C00 find similar trends in the type and environment
dependence of the GLF. The characteristic density $n^*$ varies
with both type and environment, but there is no obvious trend.
The faint end slope $\al$ systematically steepens from early-type
to late-type galaxies and from the field to denser environments,
leading to an increase in the dwarf-to-giant ratio with local
galaxy density. The fact that both studies reveal similar trends
using different galaxy classification techniques suggests that the
trends are robust. Zabludoff \& Mulchaey (2000) detect a similar
variation in the dwarf-to-giant ratio with environment using a
completely different sample of galaxies.

A drawback of the LCRS for our analysis is that the sample is drawn
from the nearby universe ($\avg{z} \simeq 0.1$, Shectman et al.\
1996), while most lens galaxies are found at redshifts between 0.3
and 1. There are, however, several independent studies of redshift
evolution in the GLF (e.g., Lilly et al.\ 1995; Lin et al.\ 1999).
We cannot use these studies directly because they are too small to
permit environment classification, but we can apply the inferred
redshift trends. We adopt as our main hypothesis that the only
evolution is passive luminosity evolution, but we also examine how
number density evolution would affect our conclusions.

\subsection{Lens model}

The gravitational lens model specifies what kind of lensed images
can be produced by a galaxy of a given luminosity, type, and
environment. We adopt the standard singular isothermal sphere (SIS)
lens model because of its analytic simplicity, and because it is
consistent with stellar dynamical models (e.g., Rix et al.\ 1997b),
X-ray galaxies (e.g., Fabbiano 1989), models of individual lens
systems (e.g., Kochanek 1995; Grogin \& Narayan 1996), and lens
statistics (e.g., Maoz \& Rix 1993; Kochanek 1993a, 1996a). The
image separation $\th$ produced by an SIS lens is independent of
the impact parameter of the source relative to the lens and is
given by (e.g., Schneider et al.\ 1992)
\begin{equation}
  \th = 8\pi\, \left({\s \over c}\right)^2\, {D_{ls} \over D_{os}}\ ,
  \label{eq:th}
\end{equation}
where $\s$ is the velocity dispersion of the lens galaxy, and
$D_{os}$ and $D_{ls}$ are proper motion distances from the observer
to the source and from the lens to the source, respectively. (The
distance ratio is the same if the two distances are taken to be
angular diameter distances.) The cross section for multiple imaging
(in angular units) is
\begin{equation}
  \A = {\pi \over 4}\,\th^2\,. \label{eq:A}
\end{equation}
It is convenient to introduce a characteristic angular scale for
lensing,
\begin{equation}
  \th^* = 8\pi\, \left({\s^* \over c}\right)^2 ,
  \label{eq:thstar}
\end{equation}
which is the image separation produced by an $L^*$ lens galaxy
(with velocity dispersion $\s^*$) for a source at infinity.

The SIS lens model omits three features that are known to affect
lensing: ellipticity in the lens galaxy, tidal shear from objects
near the lens galaxy, and gravitational focusing (or ``convergence'')
from the environment. Statistically, ellipticity and shear mainly
affect the relative numbers of two- and four-image lenses (e.g.,
Keeton et al.\ 1997), which we do not differentiate. Convergence
from the environment can increase image separations by a few
percent up to $\sim$20\% (e.g., Bernstein \& Fischer 1999). Our use
of the SIS lens model amounts to an assumption that environmental
contributions to the lensing potential do not affect lens
statistics; thus our results will serve as the null hypothesis when
examining whether potential perturbations affect lens statistics.

To apply SIS lens models to galaxies classified by luminosity, we
must convert from luminosity to velocity dispersion using empirical
scaling relations like the Faber-Jackson relation for early-type
galaxies and the Tully-Fisher relation for late-type galaxies. Both
relations have the form
\begin{equation}
  {L \over L^*} = \left({\s \over \s^*}\right)^\gamma .
  \label{eq:FJ}
\end{equation}
For early-type galaxies, combining the Faber-Jackson relation with
gravitational lens statistics yields (Kochanek 1993a)
\begin{equation}
  \g \approx 4, \quad
  \s^*=220\pm20\ \kms, \quad
  \th^*=2\farcs79,
  \label{eq:Enorm}
\end{equation}
which also agrees with dark matter models for the stellar dynamics
of elliptical galaxies (Kochanek 1994). For spiral galaxies, we
find that combining the R-band GLF (\reftabs{B98lf}{C00lf}) with
the R-band Tully-Fisher relation of Sakai et al.\ (2000) yields
\begin{equation}
  \g=3.6\pm0.3, \quad
  \s^*=100\pm6\ \kms, \quad
  \th^*=0\farcs58.
  \label{eq:Lnorm}
\end{equation}
This is an updated version of the normalization given by Fukugita
\& Turner (1991). The normalization differences between early-type
and late-type galaxies are basically differences in mass. Early-type
galaxies tend to have more mass for a given luminosity (a higher
mass-to-light ratio) than late-type galaxies, at least within the
optical radius, so they tend to have higher velocity dispersions
and produce larger image separations.

We assume that $\s^*$ and $\g$ do not change with redshift. Under
passive luminosity evolution, the mass scale given by $\s^*$ does
not evolve. Also, high-redshift studies of the Tully-Fisher
relation (e.g., Vogt et al.\ 1996; Rix et al.\ 1997a) and of the
Fundamental Plane of elliptical galaxies (e.g., van Dokkum \&
Franx 1996; Kelson et al.\ 1997; van Dokkum et al.\ 1998;
Kochanek et al.\ 2000a) suggest that $\g$ does not evolve. Even
if the no-evolution assumption is not entirely justified, Mao
(1991), Mao \& Kochanek (1994), and Rix et al.\ (1994) have shown
that lens statistics are affected by evolution only if there are
dramatic changes in the population of early-type galaxies at
redshifts $z < 1$.

\subsection{Statistics}

We want to compute statistical properties of both non-lens and
lens galaxies. For simplicity, in this discussion we refer to
the set of galaxies in a particular type and environment bin $i$
as galaxy population $i$. The total number density $n_i$ and
luminosity density $\rho_{L,i}$ for galaxy population $i$ are
obtained by integrating over the luminosity function,
\begin{eqnarray}
  n_i &=& \int_{\Lcut}^{\infty} \phi_i(L)\,dL
    = n^*_i\,\G{1+\al_i}{\Lcut/L^*_i}\,, \label{eq:n} \\
  \rho_{L,i} &=& \int_{\Lcut}^{\infty} L\,\phi_i(L)\,dL
    = n^*_i\,L^*_i\,\G{2+\al_i}{\Lcut/L^*_i}\,, \label{eq:rho}
\end{eqnarray}
where the integrals include only galaxies brighter than some
completeness limit $\Lcut$, and they can be evaluated in terms
of incomplete $\Gamma$ functions.

For lens galaxies, the main statistical quantity is the optical
depth for a source at redshift $z_s$ to be lensed by galaxy
population $i$ (e.g., Turner et al.\ 1984; Fukugita \& Turner 1991),
\begin{equation}
  \tau_i(z_s) = {1 \over 4\pi} \int dV \int dL\, \A_i(L,z_l,z_s)\,
    \phi_i(L)\,, \label{eq:tauint}
\end{equation}
where $\phi_i$ is the GLF for the galaxy population and
$\A_i(L,z_l,z_s)$ is the cross section for multiple imaging by a
lens of luminosity $L$ at redshift $z_l$. In order to compute the
number of lenses expected in any real survey, the optical depth
would have to be modified by a ``magnification bias'' factor to
account for lenses where the source is intrinsically fainter than
the flux limit but magnified above the threshold (e.g., Turner
1980; Turner et al.\ 1984). For the SIS lens model, magnification
bias simply yields a coefficient multiplying the optical depth, and
the coefficient is the same for all galaxy populations. Because we
are mainly interested in the relative number of lenses produced by
different galaxy populations, we can neglect magnification bias.

If there is no redshift evolution in the GLF other than passive
luminosity evolution, the optical depth and other related quantities
can be computed analytically. The following results were originally
given by Gott, Park \& Lee (1989), Fukugita \& Turner (1991), Kaiser
(1991), and Kochanek (1993b); we have modified them to introduce an
explicit luminosity cut at the completeness limit $\Lcut$. The
total optical depth is
\begin{eqnarray}
  \tau_i(z_s) &=& \tau^*_i\,f(z_s)\,\G{1+\al_i+4\g_i^{-1}}{\Lcut/L^*_i}\,,
    \label{eq:tau} \\
  \mbox{where}\quad
  \tau^*_i &=& 16\pi^3\,r_H^3\,n^*_i\,\left({\s^*_i \over c}\right)^4\,, \\
  f(z_s) &=& {1 \over r_H^3} \int_0^{D_{os}}
    { (D_{ol} D_{ls} / D_{os})^2 \over
    (1+\Omega_k D_{ol}^2/r_H^2)^{1/2} }\,dD_{ol}\,,
    \label{eq:fz}
\end{eqnarray}
where $D_{ol}$ is the proper motion distance from the observer to
the lens, $r_H = c/H_0$ is the Hubble distance, and $\Omega_k = 1 -
\Omega_M - \Omega_\Lambda$ is the curvature density. Writing the
optical depth as in \refeq{tau} emphasizes how the various
dependences separate. The characteristic optical depth $\tau^*_i$
contains the characteristic number density $n^*_i$ and mass scale
$\s^*_i$ for the galaxy population. The $\Gamma$ function contains
the shape information from the GLF (the faint end slope $\al_i$)
and the $L$--$\s$ scaling relation (the Faber-Jackson or
Tully-Fisher slope $\g_i$). Finally, the dimensionless factor
$f(z_s)$ contains all of the dependence on cosmology and on the
source redshift; for any flat universe ($\Omega_k=0$), $f(z_s) =
D_{os}^3 / (30 r_H^3)$ (Gott et al.\ 1989). When considering a
population of sources, the cosmology factor $f$ would also include
an integral over the source redshift distribution.

We characterize the distribution of lenses using the differential
optical depth $d\tau_i/d\th$, which is equivalent (up to a
normalization factor that depends on the cosmology and the source
redshift distribution) to a probability distribution for the image
separation $\th$. Hence we refer to $d\tau_i/d\th$ as the image
separation distribution. It can be evaluated analytically for any
flat cosmology,
\newcommand\that{\hat\th}
\begin{eqnarray}
  {d\tau_i \over d\th} &=& 30\,\tau^*_i\,f(z_s)\,{\that^2 \over \th^*_i}\,
    \left( \Gamma_2 - 2\,\that\,\Gamma_4 + \that^2\,\Gamma_6 \right)\,,
    \label{eq:dtau} \\
  \Gamma_n &\equiv& \Gamma\left[1+\al_i-n \g_i^{-1} \,;\,
    \max(\that^{\g_i/2},\Lcut/L^*_i) \right]\,,
\end{eqnarray}
where $\that=\th/\th^*_i$. Moments of this distribution can also
be evaluated analytically for any flat cosmology,
\begin{eqnarray}
  \avg{\th}_i &=& {\th^*_i \over 2}\
    {\G{1+\al_i+6\g_i^{-1}}{\Lcut/L^*_i} \over
     \G{1+\al_i+4\g_i^{-1}}{\Lcut/L^*_i}}\ ,
    \label{eq:thavg} \\
  \avg{\th^2}_i &=& {2\left(\th^*_i\right)^2 \over 7}\
    {\G{1+\al_i+8\g_i^{-1}}{\Lcut/L^*_i} \over
     \G{1+\al_i+4\g_i^{-1}}{\Lcut/L^*_i}}\ ,
\end{eqnarray}
where $\avg{\cdots}_i$ denotes an average over galaxy population
$i$. The mean image separation is simply $\avg{\th}_i$, while the
standard deviation is $[\avg{\th^2}_i-\avg{\th}_i^2]^{1/2}$.

All of these results apply for a single galaxy population in a flat
cosmology. For the generalization to arbitrary cosmologies, see
Kochanek (1993b). For the generalization to multiple galaxy
populations, the linearity of the optical depth means that we can
simply sum over populations,
\begin{eqnarray}
  \tau_\tot &=& \sum_i \tau_i\ , \label{eq:tautot} \\
  {d\tau_\tot \over d\th} &=& \sum_i {d\tau_i \over d\th}\ .
    \label{eq:dtautot}
\end{eqnarray}
When computing moments of the image separation distribution, we
must be careful to weight the populations correctly,
\begin{equation}
  \avg{\th}_\tot = {\sum_i \tau_i\,\avg{\th}_i \over
    \sum_i \tau_i}\ , \label{eq:avgtot}
\end{equation}
and similarly for $\avg{\th^2}_\tot$.

Finally, all of these results apply to a galaxy population with no
redshift evolution other than passive luminosity evolution. However,
redshift surveys suggest that there is evolution in the comoving
number density of late-type galaxies (Lilly et al.\ 1995; Lin et
al.\ 1999). Adding number density evolution $n^*_i(z)$ to the
statistical results given above is straightforward: evaluate
$\tau^*_i$ using the local number density $n^*_i(0)$ and replace
the cosmology factor $f(z_s)$ with the cosmology/evolution factor
\begin{equation}
  \fevol(z_s) = {1 \over r_H^3} \int_0^{D_{os}}
    {n^*_i(z_l) \over n^*_i(0)}\,
    { (D_{ol} D_{ls} / D_{os})^2 \over
    (1+\Omega_k D_{ol}^2/r_H^2)^{1/2} }\,dD_{ol}\,.
    \label{eq:fevol}
\end{equation}

Note that making an explicit luminosity cut at the LCRS completeness
limit $M_R = -17.5 + 5 \log h$ excludes some galaxies and therefore
yields an underestimate of the lensing optical depth. The cut is
necessary because the luminosity function for any survey is
unreliable below the survey's completeness limit. However, the cut
does not significantly affect our results, because the faint
galaxies below it are very poor lenses. To estimate the amount of
optical depth excluded by the cut, consider a GLF that matches an
observed GLF about the completeness limit, but is allowed to have
an arbitrary slope $\al_\faint$ below the completeness limit. Let
$\tau_\bright$ ($\tau_\faint$) be the lensing optical depth due to
galaxies brighter (fainter) then the completeness limit. Using the
C00 sample of NEL galaxies outside groups (see \reftab{C00lf}), we
estimate $\tau_\faint/\tau_\bright = (0.005, 0.008, 0.016)$ when
the slope below the completeness limit is $\al_\faint = (-0.5, -1.0,
-1.5)$. In other words, even if the GLF is much steeper below the
completeness limit than above the limit, the excluded optical depth
is a small fraction of the total.

\section{Analytic Trends}

Before making quantitative predictions about the effects of type
and environment in lens statistics, it is instructive to use the
analytic results from \S 2.3 to identify the general trends. There
are two familiar effects related to galaxy type, and two new
effects due to the environment.

Because lensing selects galaxies by mass and early-type galaxies
tend to be more massive than late-type galaxies, lensing has a
strong type selection in favor of early-type galaxies. \refEq{tau}
shows that the lensing optical depth is proportional to $n^*
\left(\s^*\right)^4 \Gamma(1+\al+4\g^{-1})$, so the relative number of
lenses is expected to be $N(\mbox{early}) / N(\mbox{late}) \sim
20$, using the data from \S 2. Moreover, early-type galaxies tend
to produce lenses with larger image separations than late-type
galaxies. \refEq{thavg} indicates that the mean image separation
$\avg{\th}$ scales as $\left(\s^*\right)^2$, so $\avg{\th}$ for
late-type lenses is only $\sim$20\% of that for early-type lenses,
although this simple estimate may be complicated by inclination
effects due to the thin disk in spiral galaxies (see Maller,
Flores \& Primack 1997; Wang \& Turner 1997; Keeton \& Kochanek
1998). Both of these effects are known from previous calculations
of lens statistics (e.g., Turner et al.\ 1984; Fukugita \& Turner
1991). They are also consistent with the data, as most of the more
than 50 known lenses are produced by early-type galaxies (e.g.,
Kochanek et al.\ 2000a), with only four likely cases of lensing
by a spiral galaxy (B 0218+357, Browne et al.\ 1993; B 1600+434,
Jaunsen \& Hjorth 1997; PKS 1830$-$211, Wiklind \& Combes 1996;
2237+0305, Huchra et al.\ 1985). Also, the smallest known image
separation is produced by a face-on spiral galaxy ($\th = 0\farcs33$
for B 0218+357). The type dependence of the number density of lenses
and the mean image separation leads to an environmental dependence
because of the morphology--density relation, as we show in the
next section.

\begin{figure}
\centerline{\epsfxsize=6.0in \epsfbox{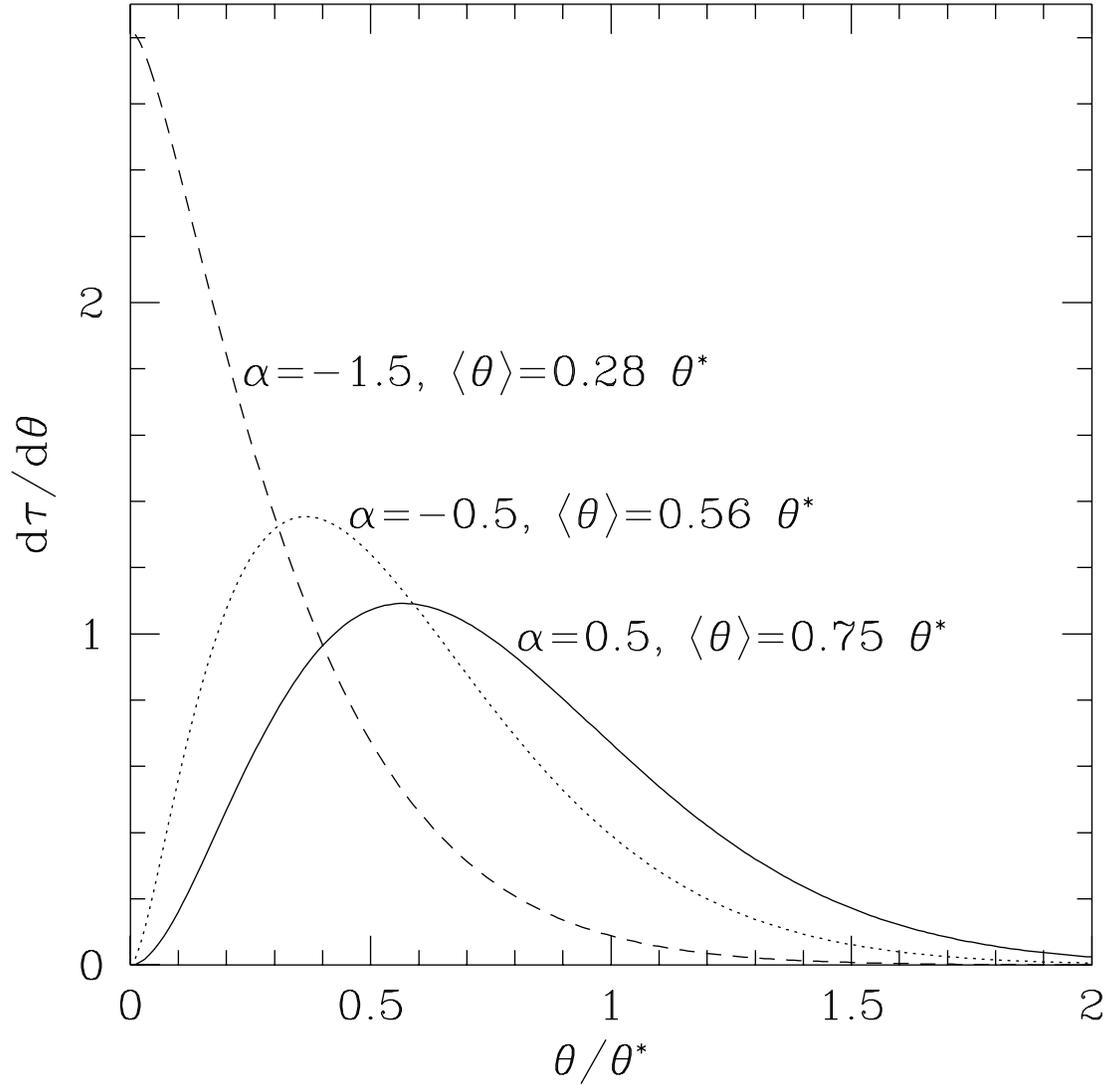}}
\caption{
Image separation distributions $d\tau/d\th$ for environments with
different values of the faint end slope $\al$ of the galaxy
luminosity function. All of the curves use a Faber-Jackson or
Tully-Fisher slope $\g=4$ and are normalized to have unit area.
The mean image separation $\avg{\th}$ for each curve is indicated.
}\label{fig:sep}
\end{figure}

The two new environmental effects arise from the environmental
dependence of the faint end slope $\al$ of the GLF. First, because
the faint end slope appears to systematically change with
environment (see \reftabs{B98lf}{C00lf}, and Zabludoff \& Mulchaey
2000), the distribution of image separations should vary with
environment. \reffig{sep} shows the image separation distribution
$d\tau/d\th$ for different values of $\al$. As $\al$ decreases, the
distribution becomes more skewed toward small separations, and
$\avg{\th}$ decreases. Physically, when $\al$ is smaller the galaxy
population has a higher ratio of dwarf galaxies to giant galaxies,
which leads to a higher fraction of small-separation lenses.

\begin{figure}
\centerline{\epsfxsize=6.0in \epsfbox{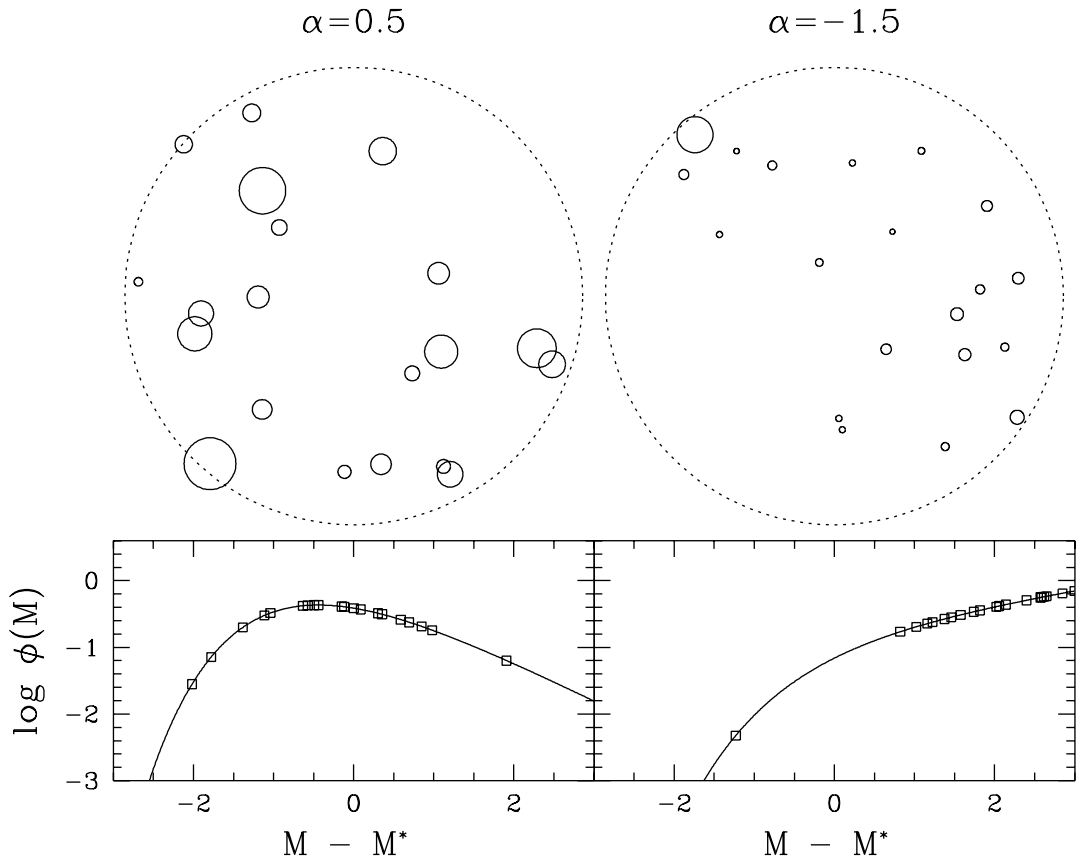}}
\caption{
Schematic representations of groups with $\al=0.5$ (left) and
$\al=-1.5$ (right), normalized to have the same number of galaxies
down to $M^*+3$. The luminosity functions are shown at the bottom.
For each group, galaxies are chosen randomly from the luminosity
function, placed randomly inside the dotted circle, and drawn with
a circle whose area is proportional to the galaxy's luminosity and
lensing optical depth and whose diameter is proportional to the
image separation. (The random placement of galaxies is not
realistic, but it is done for illustration only and does not affect
our calculations.) The points on the luminosity function indicate
the galaxies.
}\label{fig:toy1}
\end{figure}

Second, changes in $\al$ with environment affect the numbers of
lens and non-lens galaxies differently. Consider \reffig{toy1},
which shows schematic representations of groups with $\al=0.5$ and
$-1.5$, normalized to have the same number of galaxies down to
$M^*+3$. If we pick a galaxy at random, it is equally likely to
come from either group. However, if we pick a lens galaxy at random,
it is far more likely to come from the $\al=0.5$ group than the
$\al=-1.5$ group, because the latter group is dominated by dwarf
galaxies. Considering other toy models, such as groups normalized
to have the same lensing optical depth, leads to a general
conclusion: other things being equal (i.e., in the absence of a
morphology--density relation), lens galaxies are less likely than
non-lens galaxies to be found in environments with a large
dwarf-to-giant ratio. The effect is probably weaker in observed
galaxy populations than in our toy models with their rather
extreme values of $\al$.

\section{Quantitative Results}

We now evaluate lens statistics using empirical GLFs to quantify
the effects of the morphology--density relation and the
dwarf-to-giant ratio on lens statistics. In \S 4.1 we present
results using the full set of type and environment bins defined
by B98 and C00, while in \S 4.2 we use coarser binning with only
two types (early- and late-type galaxies) and three environments
(the field, poor groups, and rich clusters). In \S 4.3 we compare
our results to the data. In \S\S 4.1--4.3 we assume no evolution
in the comoving number density of galaxies with redshift, but in
\S 4.4 we discuss how evolution would affect our results. Finally,
in \S 4.5 we discuss the effects of possible incompletenesses in
the Las Campanas Redshift Survey.

\subsection{Fine binning}

\begin{figure}
\centerline{\epsfxsize=6.0in \epsfbox{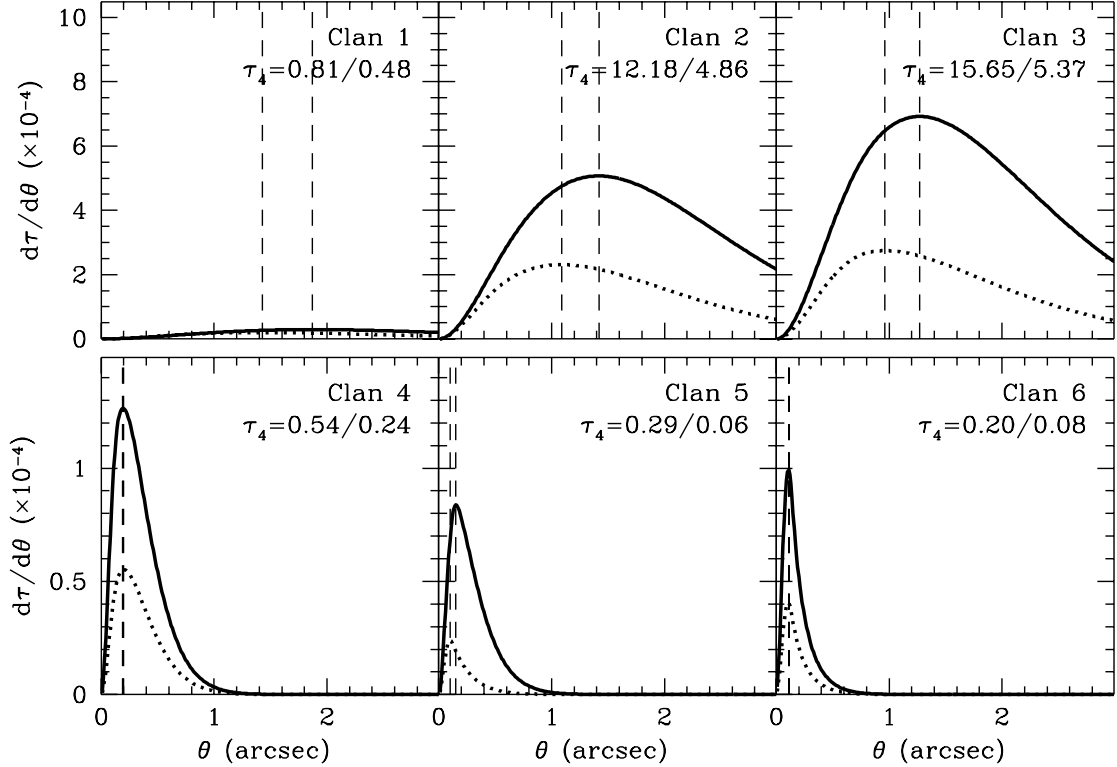}}
\caption{
Image separation distributions for the six clans and two
environments defined by B98. Results are shown for a source
at redshift $z_s = 2$ in a cosmology with $\Omega_M = 0.3$ and
$\Omega_\Lambda = 0.7$. Each panel contains a single clan; the
solid and dotted curves denote low and high density environments,
respectively. Note that the vertical scales differ between the
two rows. The vertical dashed lines indicate the peaks of the
distributions. The two values of $\tau_4$ in each panel give
the optical depth (in units of $10^{-4}$) for low and high
density environments, respectively. We treat Clans 1--3 as
early-type galaxies and Clans 4--6 as late-type galaxies; if
we were to reclassify Clan 3 as late-type galaxies, the results
would look very similar to the Clan 4 panel.
}\label{fig:B98sep}
\end{figure}

\reffigs{B98sep}{C00sep} show the image separation distributions
$d\tau_i/d\th$ for the type and environment bins defined by B98 and
C00, respectively. The distributions are computed assuming a source
redshift $z_s = 2$ and a cosmology with $\Omega_M = 0.3$ and
$\Omega_\Lambda = 0.7$. If we were to change the cosmology or the
source redshift, or even to allow a distribution of source
redshifts, the only effect would be to change $d\tau_i/d\th$ by a
multiplicative factor that is identical for all type and environment
bins (see eq.~\ref{eq:tau}). The relative contributions of the
different bins to the lensing optical depth are independent of the
cosmology and source redshift distribution.

The results illustrate the trends identified in \S 3. First, for
galaxies in a particular environment, both the amplitude and the
peak location of the distributions are larger for early-type
galaxies than for late-type galaxies; relative to early-type
galaxies, late-type galaxies have a much smaller lensing optical
depth and produce smaller image separations. Second, for galaxies
of a particular type, the lens distributions in denser environments
are more dominated by small image separations. This effect occurs
because denser environments tend to have a steeper GLF and hence a
higher dwarf-to-giant ratio (B98; C00; Zabludoff \& Mulchaey 2000),
which means a higher fraction of galaxies that produce small
separation lenses. These trends are seen in both the B98 and C00
samples, which suggests that they are not overly sensitive to
details of the type and environment classification schemes.

\subsection{Coarse binning}

The type classification of B98 and the environment classification
of C00 both yield finer differentiation than we seek for
characterizing the type and environment distributions of lens
galaxies. We confine the B98 sample to two type bins by defining an
early-type sample using Clans 1--3 and a late-type sample with
Clans 4--6. We reclassify the C00 sample into three environment
bins: the field, poor groups, and rich clusters. Studies of groups
suggest that many systems with velocity dispersions $\s \lesssim
200$ \kms\ are either spurious or similar to the Local Group, which
is bound but not yet virialized (Diaferio et al.\ 1993; Zaritsky
1994; Zabludoff \& Mulchaey 1998). Hence, to be conservative, we
define a ``field'' bin that contains both galaxies not in groups
and galaxies in systems with $\s < 200$ \kms, a ``group'' bin that
contains galaxies in systems with $200 < \s < 500$ \kms, and a
``cluster'' bin that contains galaxies in systems with $\s > 500$
\kms. As we demonstrate below, our main conclusions are robust to
the effects of changing the definitions of the coarse bins. Note
that we compute statistical quantities for the full set of type and
environment bins, and then combine the results into the coarse bins
using eqs.~(\ref{eq:tautot})--(\ref{eq:avgtot}); we do not try to
define a GLF for each coarse bin.

\begin{figure}
\centerline{
  \epsfxsize=3.0in \epsfbox{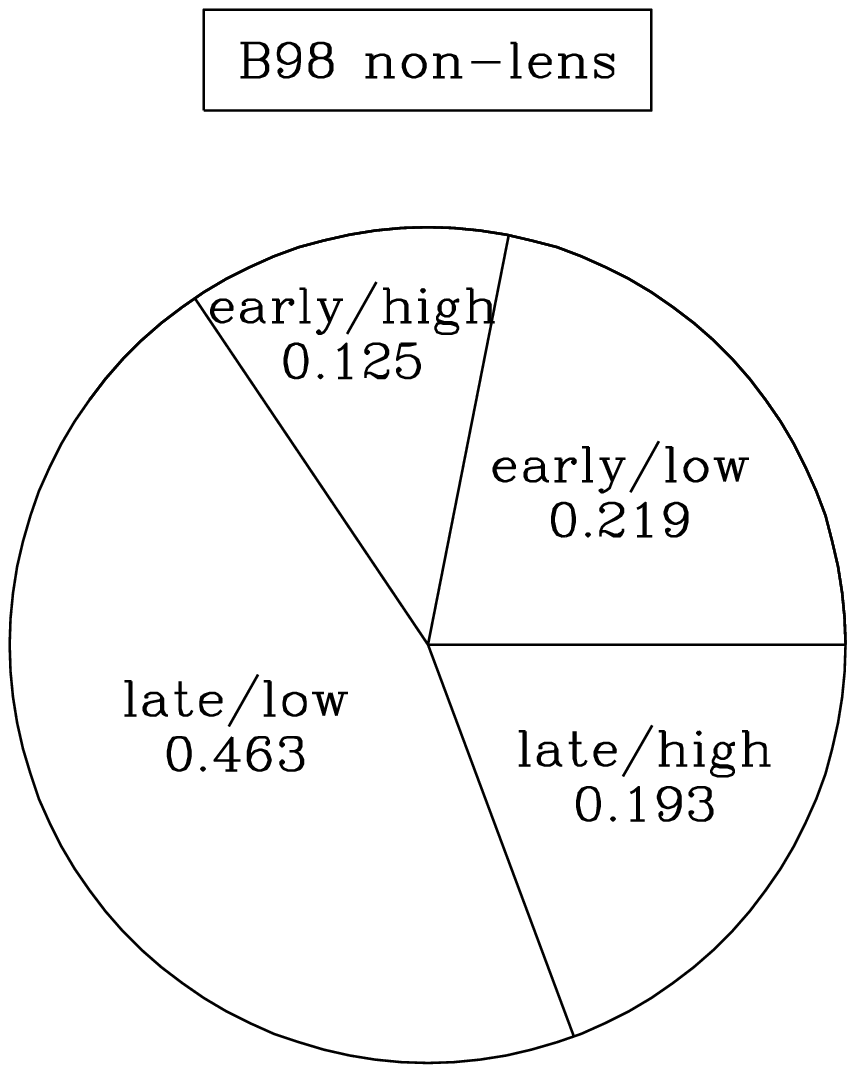}
  \epsfxsize=3.0in \epsfbox{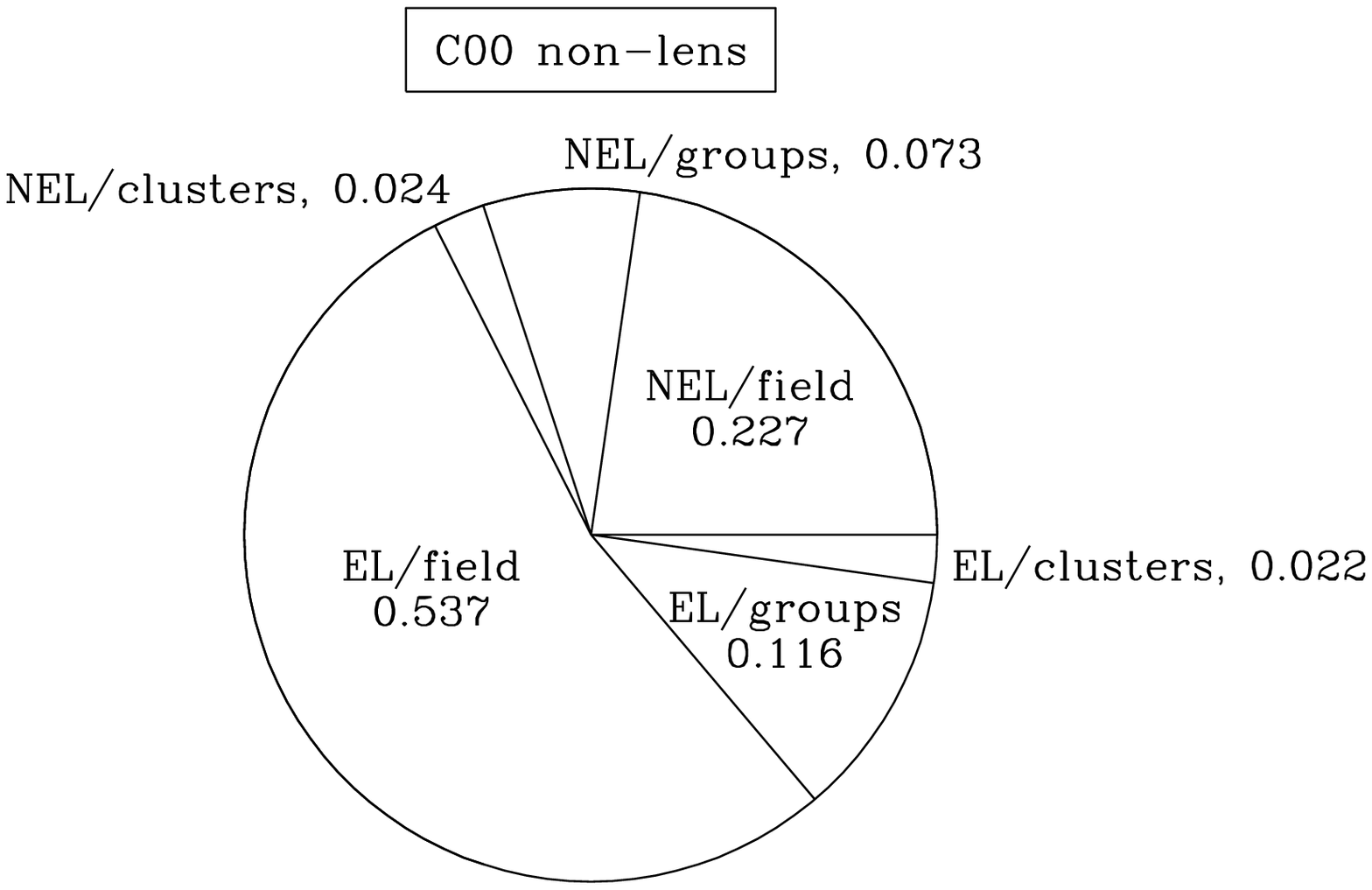}
}
\centerline{
  \epsfxsize=3.0in \epsfbox{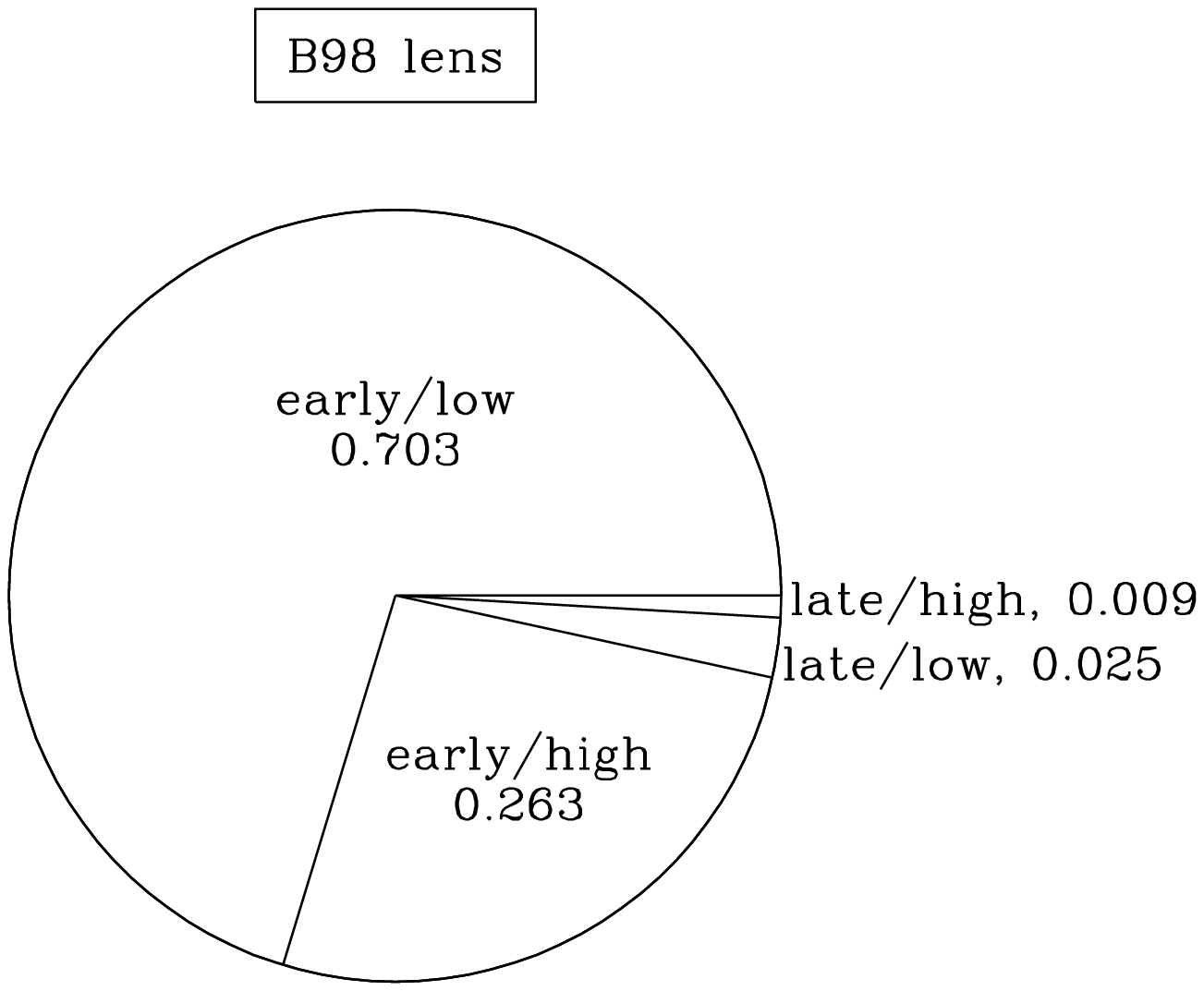}
  \epsfxsize=3.0in \epsfbox{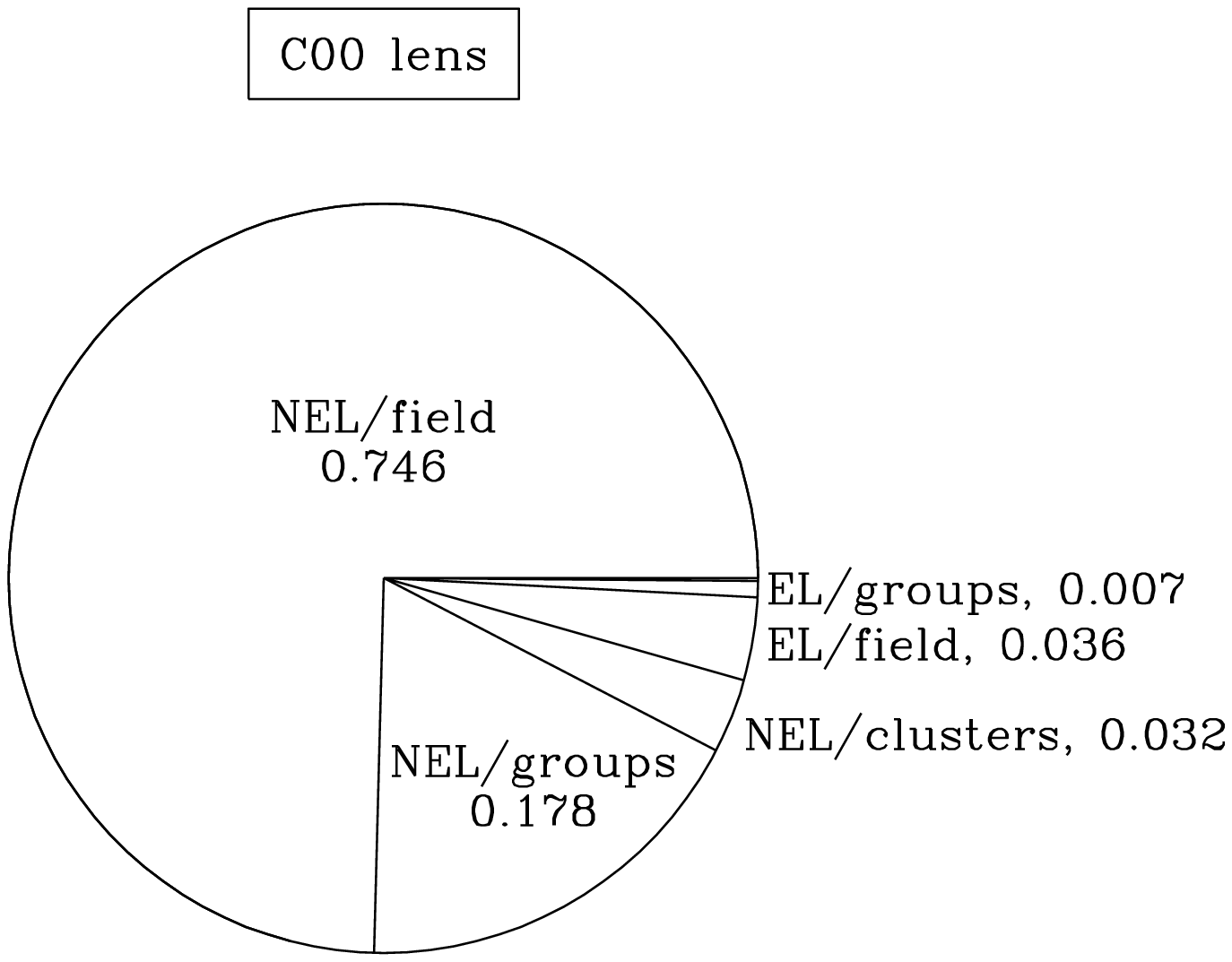}
}
\caption{
Pie charts giving the relative contributions of the different
type and environment bins to the set of non-lens galaxies (top)
and the set of lens galaxies (bottom). Results are shown for
the B98 (left) and C00 (right) samples defined in the text.
For the C00 lens sample, the EL/clusters bin has a fraction
0.001 and is too small to be seen.
}\label{fig:pie}
\end{figure}

\reffig{pie} shows the relative contributions of the coarse type
and environment bins to the sets of non-lens and lens galaxies.
We emphasize that the results for non-lens galaxies are empirical
results from observed galaxy samples, while the results for lens
galaxies are predictions based on the SIS lens model. The most
dramatic result is the familiar type difference between non-lens
and lens galaxies. Late-type or EL galaxies account for 66\% and
68\% of non-lens galaxies in the B98 and C00 samples (respectively),
but only 4\% of lens galaxies in either sample.\footnote{Changing
the type classification of the B98 Clan 3 and 4 galaxies would of
course change the type fractions. Reclassifying Clan 3 galaxies
as late-type would increase the fraction of late-type non-lens
galaxies to 85\% and the fraction of late-type lens galaxies to
11\%. Conversely, reclassifying Clan 4 galaxies as early-type
would reduce the non-lens late-type fraction to 43\% and the lens
late-type fraction to 1\%. The effects are smaller for lens
galaxies than for non-lens galaxies because late-type galaxies
contribute a small fraction of lenses to begin with.} Again,
these results reiterate the well-known fact that lensing selects
galaxies by mass and overwhelmingly favors early-type galaxies.

The environment fractions are given in \reftab{envfrac}. The
morphology--density relation is evident in the environment
fractions for non-lens galaxies; it can also be seen in the NEL
fraction, which for the C00 sample is $(0.30, 0.39, 0.52)$ in
(the field, groups, clusters). Acting alone, the
morphology--density relation would imply that lens galaxies
should be more likely than non-lens galaxies to lie in dense
environments. But lensing is also affected by changes in the
dwarf-to-giant ratio with environment. It turns out that the two
effects nearly cancel, and the environment fractions are very
similar for non-lens and lens galaxies. In fact, the changing
dwarf-to-giant ratio is the slightly stronger effect. Hence
lens galaxies are slightly less likely than non-lens galaxies
to lie in dense environments, although the differences are small
and may be hard to detect. The similarity between the results
computed with the B98 and C00 samples again suggests that
systematic differences in classifying galaxy types and environments
have little effect on the relative contributions of different
environments to lens statistics.

Quantitatively, we predict that 22--27\% of lens galaxies are
in high density, bound systems like groups and clusters --- and
that the majority of these are in poor groups rather than rich
clusters.  These results are consistent with known lenses, where
the sample of more than 50 lenses includes four lens galaxies
in clusters and three in confirmed groups, but most lens galaxy
environments remain undetermined. Our predictions yield a lower
limit on the fraction of lens systems where the lensing potential
is probably perturbed by elements other than the lens galaxy. If
the lens galaxy lies in a bound group or cluster, it is very
likely that other group galaxies and an extended dark matter halo
contribute shear and/or convergence to the lensing potential.
Even if the lens is not in a bound system, however, there may
still be neighboring galaxies or structures along the line of
sight that noticeably perturb the lensing potential.

The other trend identified in \S 3 is that the mean image
separation depends on the faint end slope $\al$, which varies with
environment. \reftab{mean} gives the predicted mean and standard
deviation of the image separations for lenses in different
environments. We predict that the mean separation for lenses in
groups is smaller than the mean for lenses in the field, {\it under
the assumption that the environment does not significantly affect
the image separation\/} (see \S 2.2). The difference in the means
is considerably smaller than the standard deviation, so it would
take 20--40 group lenses to detect the difference even at the $1\s$
level (assuming that for $N$ items with standard deviation $\s$ the
uncertainty in the mean is $\s/\sqrt{N}$). Given good statistics,
however, comparing the mean separations for group lenses and field
lenses would provide an excellent test of environmental
contributions to lensing. Specifically, an extended dark halo in a
group containing a lens can provide extra gravitational focusing
that increases the image separation beyond that produced by the
lens galaxy alone (e.g., Falco et al.\ 1985). Thus if observed
lenses reveal $\avg{\th}(\mbox{groups}) > \avg{\th}(\mbox{field})$
contrary to our prediction, it would provide direct evidence that
lenses in groups are significantly affected by group dark halos.

\subsection{Comparison with data}

\begin{figure}
\centerline{\epsfxsize=6.0in \epsfbox{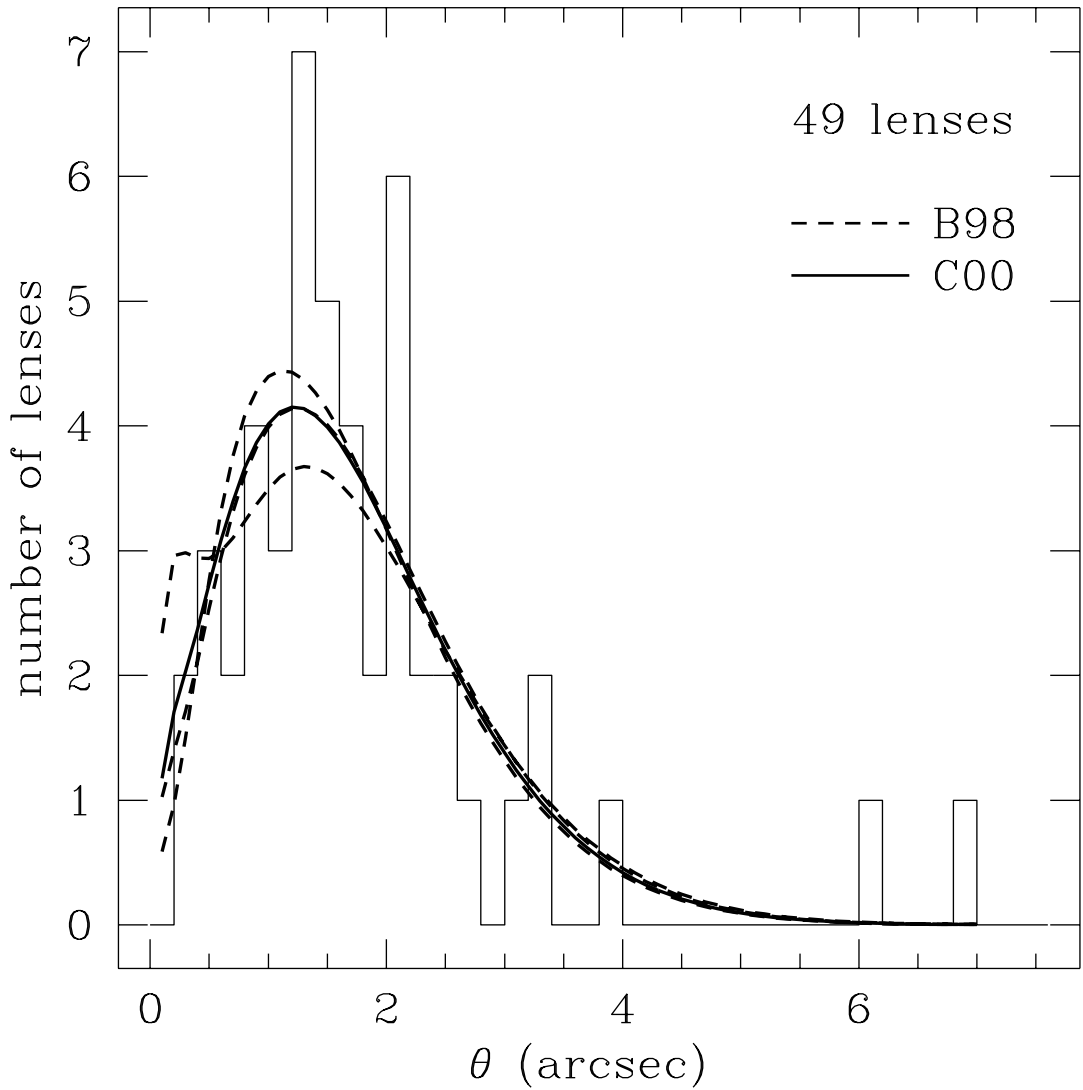}}
\caption{
The histogram shows the distribution of image separations for
49 known lenses, in bins of width $0\farcs2$. The solid curve
shows the net distribution for the C00 sample (a sum of the
curves shown in Figure 4). The three dashed curves show the
net distributions for three different models based on the B98
sample. The middle curve represents the model discussed in
the text, in which Clans 1--3 are treated as early-type
galaxies and Clans 4--6 as late-type galaxies (a sum of the
curves in Figure 3). In the higher-peaked curve Clans 1--4
are treated as early-type galaxies, while in the lower-peaked
curve only Clans 1--2 are treated as early-type galaxies. All
model curves are normalized to 49 lenses, but their shapes are
not tuned.
}\label{fig:data}
\end{figure}

\reffig{data} shows our predictions for the total image separation
distribution compared with the data for 49 known lenses from
Kochanek et al.\ (2000b). Such comparisons are sensitive to the
source redshift distribution and to cosmological parameters (e.g.,
Kochanek 1996ab; Falco et al.\ 1998; Helbig et al.\ 1999; and
references therein). With the SIS lens model, these dependences
appear only in a normalization factor that does not affect the
shape of the separation distribution (see eq.~\ref{eq:tau}), so we
avoid complications by normalizing the theoretical curves to 49
total lenses.

Ideally, comparisons between theory and data include a careful
account of selection biases, but that is impossible here because
\reffig{data} includes lenses from many different surveys as well
as serendipitous discoveries. Thus it is somewhat surprising to
find that the theoretical curves agree quite well with the data. A
K--S test (e.g., Press et al.\ 1992) cannot distinguish between the
data and any of the models. The agreement has little to do with the
changes in galaxy populations with environment. Using the
environment-independent LCRS GLFs given by Lin et al.\ (1986) yields
a separation distribution very similar to the curves shown in
\reffig{data}. The agreement is also insensitive to environmental
perturbations of the lensing potential, which our models neglect.
The largest known convergence from a group or cluster enclosing
a lens increases the image separation by $\sim$20\% (see Bernstein
\& Fischer 1999; Romanowsky \& Kochanek 1999). Randomly choosing
25\% of the lenses in \reffig{data} and increasing their image
separations by 20\% produces a distribution that is statistically
indistinguishable from the original distribution. In other words,
examining the image separation distribution without knowing the
environments of the lenses is a poor way to test whether environments
affect the lensing potential (except in the tail of the distribution,
as explained below). A better test is to determine the environments
and then compare lenses in groups with lenses in the field, as
discussed in \S 4.2.

Nevertheless, \reffig{data} does offer three conclusions. First,
regardless of the selection effects that are at work, it appears
that no particular class of lenses is substantially missing. For
example, even though finite resolution might bias surveys against
finding small-separation lenses, it appears that these lenses are
not significantly underrepresented. Second, \reffig{data} provides
a reassuring consistency check: our addition of population effects
to lens statistics has not degraded the agreement between theory
and data. It has not substantially improved the agreement, either,
because the image separation distribution is not very sensitive to
environmental effects for the reasons just cited.

The third point is that there are two lenses with $\th > 6\arcsec$,
RXJ 0921+4528 and Q 0957+561, that lie in the tail of the image
separation distribution. The C00 model shown in \reffig{data}
predicts only 0.04 lenses with $\th>6\arcsec$ and 1.39 lenses with
$\th>4\arcsec$. Part of the explanation for these two unlikely
lenses is that we have neglected environmental contributions to
the lensing potential. In Q 0957+561 the lens galaxy is the
brightest galaxy in a $\s \sim 700$ \kms\ cluster, and the
convergence from the cluster is thought to contribute $\sim$20\%
of the image separation (e.g., Bernstein \& Fischer 1999; Romanowsky
\& Kochanek 1999). In RXJ 0921+4528, the lens galaxy appears to be
in an X-ray cluster that may likewise contribute to the large
separation (Mu\~noz et al.\ 2000). These lenses suggest another
test of our null hypothesis that potential perturbations do not
affect lens statistics. They already indicate that convergence
from the environment can be important when the lens lies in a
cluster. As the data improve, it will be possible to extend this
result to lower mass environments like groups and determine the
importance of including environmental contributions to the lens
model in applications of lens statistics.

\subsection{Evolution effects}

Our results so far have been obtained under the assumption that the
comoving number density of galaxies does not change with redshift.
However, redshift surveys (e.g., Lilly et al.\ 1995; Lin et al.\
1999) suggest that there is number density evolution, at least in
the population of late-type galaxies. (The surveys also imply
evolution in the luminosity density, but lensing is insensitive to
the component of luminosity density evolution that is due to passive
luminosity evolution; see \S 2.) To quantify number density evolution,
Lin et al.\ (1999) introduce the parameter $P$ defined by
\begin{equation}
  n(z) = n(0)\,10^{0.4 P z} . \label{eq:nevol}
\end{equation}
In the redshift range $0.12 < z < 0.55$, Lin et al.\ (1999) find
$P \approx 3\pm1$ (1$\s$) for late-type galaxies and essentially
no number evolution in early-type galaxies.

We can include the effects of number evolution in the lens
statistics by replacing the cosmology factor $f(z_s)$ in
\refeq{tau} with the cosmology/evolution factor $\fevol(z_s)$ from
\refeq{fevol}, which is equivalent to multiplying the optical depth
by an evolution factor $\fevol(z_s) / f(z_s)$. \reftab{evol} shows
that this factor increases strongly with both the number evolution
parameter $P$ and the source redshift $z_s$. With higher $P$ the
evolution is more rapid, and with higher $z_s$ the increase in
number density between the observer and the source is larger.
Strong number evolution ($P \sim 3$) can increase the optical depth
for late-type galaxies to lens a distant source ($z_s \sim 3$) by
more than an order of magnitude. The effect is unlikely to be this
strong, however, for two reasons. First, fewer than 10\% of lenses
have sources beyond $z_s = 3$. Second, the strong evolution found
by Lin et al.\ (1999) was measured in the redshift range $0.12 < z
< 0.55$; there are no current grounds for extrapolation to $z \sim
1$ or beyond, especially using an exponential function that may
over-estimate the evolution at high redshifts.

Consider a rather dramatic number evolution factor of $\fevol(z_s)
/ f(z_s)=10$ for late-type galaxies and no number evolution for
early-type galaxies. Using the C00 luminosity functions, we then
find that late-type galaxies account for 32\% of lens galaxies ---
the rapid number evolution in late-type galaxies simply increases
the number of late-type lens galaxies. The mean image separations
correspondingly decrease: $\avg{\th} = (1\farcs33, 1\farcs29,
1\farcs03)$ for (field, group, cluster) lenses. Nevertheless,
the environment fractions are largely unchanged: 79\% of lens
galaxies in the field, 17\% in groups, and 3\% in clusters.
(Compare with \reftabs{envfrac}{mean} for the results without
number evolution.) The environment distributions for early- and
late-type lens galaxies are simply not very different, so changing
the relative fraction of lens galaxy types does not significantly
change the net distribution of lens galaxy environments.

\subsection{Incompleteness effects}

All galaxy redshift surveys are incomplete with either explicit
or implicit magnitude and surface brightness limits. Lin et al.\
(1996) estimate a completeness limit of $M_R = -17.5 + 5 \log h$
for the LCRS, and we have restricted our analysis to galaxies
brighter than this limit. We now ask whether any remaining
incompletenesses brighter  than this limit could affect the lens
statistics.

From a direct comparison of the R-band LCRS with the B-band CfA2
redshift survey, Huchra (1999) argues that the LCRS underestimates
the number of galaxies with $M_R \lesssim -16 + 5 \log h$ by a
factor of $\sim$4 and that most of the ``missing'' galaxies are
low surface brightness, late-type, emission line galaxies. Although
most of the suggested incompleteness occurs at magnitudes fainter
than we consider (and fainter than the LCRS completeness limit)
and is derived from a perilous comparison of surveys in different
bandpasses, we use it to estimate the effects of incompleteness on
our analysis.

In \S 2.3 we argue that even if the GLF below the completeness
limit is much steeper than above the limit, the optical depth from
galaxies below the limit is a tiny fraction of the total. Hence
the behavior of the GLF below the completeness limit has essentially
no effect on our results. To check our results further, we consider
the effects if the LCRS underestimates the number of galaxies even
at magnitudes brighter than the completeness limit. We increase
the number of galaxies in the $\Delta m = 1$ mag bin above the
LCRS completeness limit by the factor of $\sim$4 estimated by
Huchra (1999). Our approach is conservative, because we apply the
changes at magnitudes significantly brighter than where Huchra
(1999) proposes most of the incompleteness. If the incompleteness
applies to late-type galaxies in all environments, we find that it
changes the lens environment fractions by $<$0.001 compared with
\reftab{envfrac}, and it decreases the mean image separations by
$\sim$0\farcs02 compared with \reftab{mean}. If the incompleteness
applies only to late-type galaxies in the field (because of the
morphology--density relation), it changes the lens environment
fractions by $\sim$0.002 and the mean image separation in the
field by $\sim$0\farcs02. In other words, because it mainly
applies to faint galaxies that are poor lenses, incompleteness
has a negligible effect on our results.

\section{Conclusions}

We have studied how changes in galaxy populations with environment
affect gravitational lens statistics, quantifying the results with
data from the Las Campanas Redshift Survey (Shectman et al.\ 1996).
Two effects determine how the distribution of lens galaxy
environments differs from the distribution of normal galaxy
environments. First, the likelihood that lens galaxies lie in
groups and clusters is enhanced by the morphology--density relation
(e.g., Dressler 1980) and the fact that lensing tends to select
massive early-type galaxies. Second, it is diminished by the
systematic increase in the fraction of dwarf galaxies relative
to giant galaxies in dense environments (e.g., Bromley et al.\
1998b; Zabludoff \& Mulchaey 2000; Christlein 2000). The two
effects nearly cancel, and the fraction of gravitational lens
galaxies in a particular environment is very similar to the
fraction of non-lens galaxies found in that environment. As a
second order effect, lens galaxies are slightly less likely than
non-lens galaxies to be found in poor groups and rich clusters.
Quantitatively, we expect $\sim$20\% of lens galaxies to be found
in poor groups (defined as systems with velocity dispersions in the
range $200 < \s < 500$ \kms), and another $\sim$3\% of lens
galaxies in rich clusters (defined as systems with $\s > 500$
\kms).

Thus we predict that for $\gtrsim$25\% of lenses the lensing
potential may include significant contributions from objects
other than the lens galaxy. This result is only a lower limit for
three reasons. First, bound groups may exist below the velocity
dispersion threshold of 200 \kms\ that we have imposed. Second,
a lens that does not lie in a group or cluster can still be
perturbed by neighboring galaxies. Finally, lensed images may
be perturbed by large-scale structure along the line of sight.
External perturbations have been included in models of some
individual lenses (e.g., Grogin \& Narayan 1996; Schechter et al.\
1997; Leh\'ar et al.\ 2000), but they need to be added to
applications of lens statistics such as limits on $\Omega_M$ and
$\Omega_\Lambda$ (e.g., Kochanek 1996a; Falco et al.\ 1998; Helbig
et al.\ 1999). To do so, the statistical analysis of environmental
shear by Keeton et al.\ (1997) must be updated with knowledge of
the spatial distribution of galaxies and dark matter in poor
groups and rich clusters, not to mention correlations among member
galaxies and between galaxies and the diffuse dark halos in groups
and clusters. The improved analysis must also include lensing
effects that arise from the non-linear interaction between
ellipticity in the lens galaxy and shear from the environment
(see Keeton, Mao \& Witt 2000b).

Our calculation provides the starting point for the improved
analysis in two ways. First, we have assumed that environmental
perturbations of the lensing potential do not affect lens
statistics. Subsequent analyses can take our results as the null
hypothesis to test whether potential perturbations are important.
Second, we have presented empirical tests to determine whether
potential perturbations are important statistically. If they are
not important, we predict that the mean image separation for
lenses in groups is smaller than the mean separation for lenses in
the field, although it may take 20--40 group lenses to test this
prediction with significance. A contrary empirical result would
indicate the presence of extra gravitational focusing, not included
in our models, from matter in groups and clusters that is not
associated with the lens galaxy. A more immediate version of this
test comes from the existence of two large-separation lenses ($\th
> 6\arcsec$, RXJ 0921+4528 and Q 0957+561), which is far more than
predicted by our models. In fact, both lenses appear to lie in
clusters, and in Q 0957+561 the cluster is thought to contribute
$\sim$20\% of the image separation (see Bernstein \& Fischer 1999;
Romanowsky \& Kochanek 1999).

Out of the current sample of about 50 lenses, we expect about 10
lenses in groups and about two lenses in clusters. At present, the
environments of most lenses have not been determined. Four lens
galaxies appear to lie in clusters (RXJ 0911+0551, RXJ 0921+4528,
Q 0957+561, and HST 1411+5221), and another three in spectroscopically
confirmed groups (PG 1115+080, B 1422+231, and MG 0751+2716). Once
more lensing groups are found, they will provide a sample of groups
at redshifts $0.3 \lesssim z \lesssim 1$, which can be compared with
existing samples of nearby groups (e.g., Zabludoff \& Mulchaey 1998)
and of distant clusters (e.g., Stanford, Eisenhardt \& Dickinson
1998) to study galaxy evolution in different environments. The
existence of lenses in the groups will also permit direct studies
of the diffuse dark matter in poor groups of galaxies.

\acknowledgements
Acknowledgements: We thank Ben Bromley for providing data in
advance of publication, and Huan Lin for helpful discussions.
We also thank Chris Kochanek for comments on the manuscript,
and the anonymous referee for suggestions that improved the
presentation.



\newcommand\tm[1]{\tablenotemark{#1}}
\newcommand\0{\phantom{0}}

\begin{deluxetable}{ccrcccc}
\tablewidth{0pt}
\tablecaption{B98 Luminosity Function Parameters}
\tablehead{
 \colhead{Clan} & \colhead{Environment} & \colhead{$\al$} & \colhead{$M^*-5\log h$} & \colhead{$n^*$                            } \\
                &                       &                 &                         & \colhead{($10^{-3}\,h^3\,\mbox{Mpc}^{-3}$)}
}
\startdata
 1 & low  density & $ 1.10$ & $-20.06$ & $0.18$ \\
   & high density & $ 0.20$ & $-20.48$ & $0.21$ \\
\tableline
 2 & low  density & $ 0.17$ & $-20.10$ & $5.36$ \\
   & high density & $-0.39$ & $-20.39$ & $2.61$ \\
\tableline
 3 & low  density & $-0.09$ & $-19.81$ & $7.79$ \\
   & high density & $-0.58$ & $-20.00$ & $2.97$ \\
\tableline
 4 & low  density & $-0.65$ & $-19.88$ & $7.01$ \\
   & high density & $-0.61$ & $-19.78$ & $3.10$ \\
\tableline
 5 & low  density & $-1.05$ & $-19.80$ & $3.70$ \\
   & high density & $-1.61$ & $-20.39$ & $0.49$ \\
\tableline
 6 & low  density & $-1.94$ & $-20.09$ & $1.38$ \\
   & high density & $-1.93$ & $-20.14$ & $0.56$ \\
\enddata
\tablecomments{The luminosity function parameters derived by
B98 for their six spectral type ``clans'' and two environments.}
\label{tab:B98lf}
\end{deluxetable}

\begin{deluxetable}{ccrcc}
\tablewidth{0pt}
\tablecaption{C00 Luminosity Function Parameters}
\tablehead{
 \colhead{Galaxy} & \colhead{Group $\s$} & \colhead{$\al$} & \colhead{$M^*-5\log h$} & \colhead{$n^*$                            } \\
 \colhead{Type  } & \colhead{(\kms)        } &                    &                         & \colhead{($10^{-4}\,h^3\,\mbox{Mpc}^{-3}$)}
}
\startdata
 NEL
 & other & $-0.079$ & $-20.23$ & $  49.31$ \\
 &  50   & $ 0.202$ & $-20.04$ & $\0 2.58$ \\
 & 150   & $ 0.070$ & $-20.16$ & $\0 4.92$ \\
 & 250   & $-0.345$ & $-20.40$ & $\0 5.68$ \\
 & 350   & $-0.334$ & $-20.34$ & $\0 6.54$ \\
 & 450   & $-0.596$ & $-20.59$ & $\0 2.63$ \\
 & 750   & $-0.909$ & $-20.84$ & $\0 2.54$ \\
\tableline
 EL
 & other & $-0.908$ & $-20.14$ & $  63.74$ \\
 &  50   & $-0.870$ & $-20.18$ & $\0 3.23$ \\
 & 150   & $-0.958$ & $-20.30$ & $\0 4.84$ \\
 & 250   & $-0.932$ & $-20.18$ & $\0 6.09$ \\
 & 350   & $-1.304$ & $-20.49$ & $\0 3.39$ \\
 & 450   & $-0.708$ & $-20.03$ & $\0 3.02$ \\
 & 750   & $-1.025$ & $-20.17$ & $\0 2.58$ \\
\enddata
\tablecomments{The luminosity function parameters derived by
C00 for two galaxy type and seven environment bins. One
environment bin (``other'') contains all galaxies that do not lie
in groups. The remaining bins classify galaxies by the velocity
dispersion $\s$ of the group in which they reside. The five
environment bins labeled 50, 150, 250, 350, and 450 \kms\ are
100 \kms\ wide and labeled by the central value. The bin labeled
750 \kms\ includes the range $500 < \s < 1000$ \kms. Typical
uncertainties are $\sim$0.2 in $\al$ and $\sim$0.2 mag in $M^*$;
see C00 for details.
}
\label{tab:C00lf}
\end{deluxetable}

\clearpage

\begin{deluxetable}{rccccc}
\tablewidth{0pt}
\tablecaption{Environment Fractions}
\tablehead{
 & \multicolumn{2}{c}{B98 Sample}
 & \multicolumn{3}{c}{C00 Sample} \\
 \colhead{Galaxy Type}
 & \colhead{Low Density} & \colhead{High Density}
 & \colhead{Field} & \colhead{Groups} & \colhead{Clusters}
}
\startdata
Early-type / Non-lens & 0.636 & 0.364 & 0.700 & 0.226 & 0.074 \\
                 Lens & 0.728 & 0.272 & 0.781 & 0.186 & 0.033 \\
\tableline
 Late-type / Non-lens & 0.706 & 0.294 & 0.796 & 0.171 & 0.033 \\
                 Lens & 0.733 & 0.267 & 0.821 & 0.149 & 0.030 \\
\tableline
       All / Non-lens & 0.682 & 0.318 & 0.765 & 0.189 & 0.046 \\
                 Lens & 0.728 & 0.272 & 0.783 & 0.185 & 0.033 \\
\enddata
\tablecomments{The fraction of non-lens and lens galaxies of
different types that are expected in each environment, computed
for the samples defined in the text. We use Monte Carlo simulations
to estimate the statistical uncertainties in the environment
fractions due to measurement uncertainties in $\al$ and $M^*$
(given by C00) and $\s^*$ (see \S 2.2). We estimate that the
uncertainties are $\le\!(0.040,0.037,0.023)$ for non-lens
galaxies in (the field, groups, clusters), while they are
$\le$0.008 for lens galaxies in all environments.}
\tablenotetext{}{
These results are robust to changes in the definitions of the
type and environment bins. The B98 fractions are essentially
unchanged if we change the type classification of Clans 3 and/or
4. For the C00 sample, if we increase the lower velocity dispersion
threshold for groups to $\s = 300$ \kms, the (field, group, cluster)
fractions become $(0.839, 0.115, 0.046)$ for non-lens galaxies and
$(0.853, 0.114, 0.034)$ for lens galaxies.}
\label{tab:envfrac}
\end{deluxetable}

\begin{deluxetable}{cccc}
\tablewidth{300pt}
\tablecaption{Mean Image Separations}
\tablehead{
 \colhead{Sample} & \colhead{Environment} & \colhead{$\avg{\th}$ (\arcsec)} & \colhead{$\s_\th$ (\arcsec)}
}
\startdata
 B98 & Low Density  & 1.83 & 1.04 \\
     & High Density & 1.57 & 0.95 \\
\tableline
 C00 & Field    & 1.76 & 1.02 \\
     & Groups   & 1.61 & 0.96 \\
     & Clusters & 1.30 & 0.85 \\
\enddata
\tablecomments{The predicted mean $\avg{\th}$ and standard
deviation $\s_\th$ of the image separation distributions for
lenses in different environments. Monte Carlo simulations
indicate that the statistical uncertainties in $\avg{\th}$
are $\sim$0\farcs3, due almost entirely to uncertainties in
$\s^*$ (see \S 2.2). However we are more interested in the
difference $\avg{\th}(\mbox{field}) - \avg{\th}(\mbox{groups})$,
where the uncertainty is only 0\farcs09 and is due primarily
to uncertainties in $\al$. (In the difference quantity, $\s^*$
factors out into a constant scale factor.) Larger surveys
such as the Sloan Digital Sky Survey (e.g., Gunn \& Weinberg
1995) should reduce this uncertainty to $\lesssim$0\farcs04
by placing better constraints on $\al$.}
\label{tab:mean}
\end{deluxetable}

\begin{deluxetable}{rrrrr}
\tablewidth{300pt}
\tablecaption{Number Evolution Factors}
\tablehead{
 & \colhead{$z_s=1.5$} & \colhead{2.0} & \colhead{2.5} & \colhead{3.0}
}
\startdata
 $P=1.5$ & 2.51 &  3.28 &  4.23 &  5.44 \\
    2.0  & 3.53 &  5.17 &  7.52 & 11.00 \\
    2.5  & 5.05 &  8.41 & 14.06 & 23.90 \\
    3.0  & 7.35 & 14.10 & 27.56 & 55.57 \\
\enddata
\tablecomments{The number evolution factor $\fevol(z_s)/f(z_s)$
computed for different values of the source redshift $z_s$ and
the Lin et al.\ (1999) number evolution parameter $P$, assuming
$\Omega_M=0.3$ and $\Omega_\Lambda=0.7$.}
\label{tab:evol}
\end{deluxetable}


\clearpage

\begin{figure}
\centerline{\epsfxsize=6.0in \epsfbox{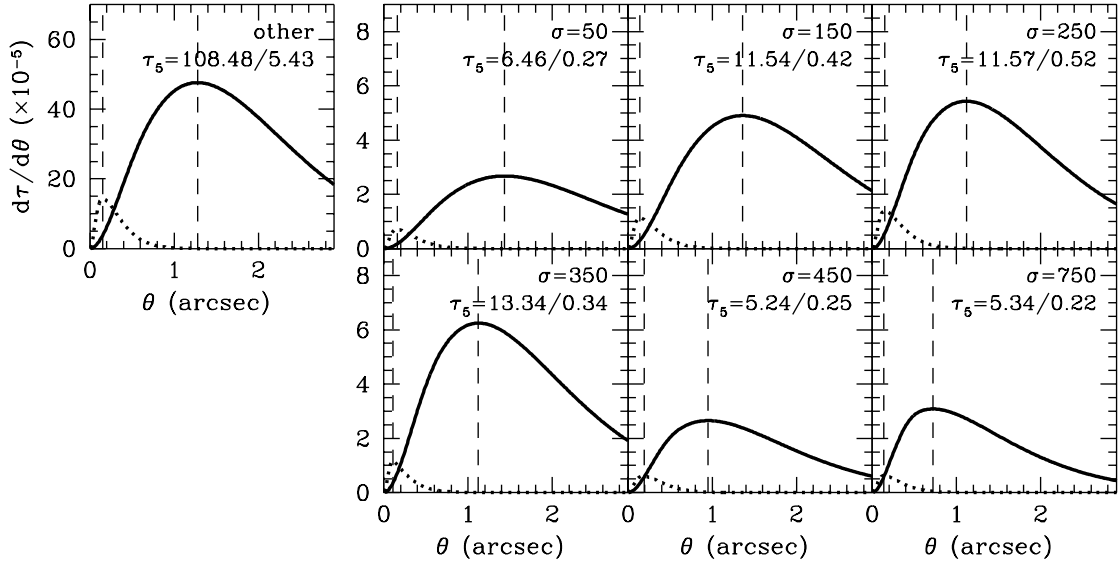}}
\caption{
Image separation distributions for the two types and seven
environments defined by C00, again for $z_s = 2$, $\Omega_M = 0.3$,
and $\Omega_\Lambda = 0.7$. Each panel contains a single
environment; the solid curves denote NEL galaxies and the dotted
curves EL galaxies. The vertical dashed lines indicate the peak
of each distribution. In each panel, the two values of $\tau_5$
give the optical depth (in units of $10^{-5}$) for NEL and EL
galaxies, respectively. Note that the vertical axis scale
for the ``other'' panel differs from that in the group panels.
For the luminosity/mass conversions, we treat NEL galaxies
as early-type galaxies and EL galaxies as late-type galaxies.
}\label{fig:C00sep}
\end{figure}

\end{document}